# ProCAVE: A Self-Adaptive, Full-Lifecycle Edge Caching Framework for Video Streaming via Predictive Bandwidth Estimation and Preference-Aware Deep Reinforcement Learning

Yeganeh Chatri[1], Behzad Akbari[1*], Foad Ghaderi[1], Pejman Goudarzi[2]
[1]*Dept. of Computer and Electronic Engineering, Tarbiat Modares University*
[2]*Dept. of Information Technology, ICT Research Institute*

***Abstract*—The growing demand for mobile video streaming requires edge delivery systems that adapt efficiently to rapid network fluctuations and diverse user preferences. Existing approaches such as FlyCache rely on reactive ABR heuristics and loosely coupled cache policies, limiting their responsiveness and coordination under real-world wireless dynamics. We propose ProCAVE(Proactive Caching with Adaptive Video Experience), a self-adaptive DRL-based framework that unifies predictive bandwidth modeling, proactive bitrate selection, and preference-aware cache control. ProCAVE employs: (i) a lightweight Transformer for short-term throughput forecasting; (ii) a PPO-driven ABR agent; and (iii) a DDPG-based continuous cache controller operating on a high-dimensional global state. Experiments using *MovieLens* preference traces and *Ghent 4G* bandwidth measurements show that ProCAVE improves byte hit rate, reduces backhaul load, and enhances QoE compared with FlyCache and other baselines. These results highlight the benefits of predictive, DRL-coordinated control for efficient and user-centric edge video delivery.**

## I. Introduction

Video streaming dominates mobile Internet traffic. Edge-assisted delivery reduces latency and backhaul load, but its effectiveness is limited by highly non-stationary wireless bandwidth and shifting user preferences. Existing edge caching frameworks, including FlyCache [1], remain reactive: they assume quasi-stable bandwidth and rely on fixed, rule-based ABR, leaving cache decisions uncoordinated from playback dynamics.

FlyCache [1] advances edge caching through a full-lifecycle framework integrating prefix placement, dynamic admission, and progressive eviction. Yet, it assumes relatively stable bandwidth and relies on external rule-based ABR algorithms (e.g., BOLA, MPC), decoupling bitrate adaptation from cache management. As a result, it cannot anticipate channel fluctuations nor coordinate cross-layer decisions in real time.

To overcome these limitations, we propose ProCAVE (Proactive Caching with Adaptive Video Experience), a predictive, DRL-based extension of FlyCache that jointly leverages (i) short-term bandwidth forecasting, (ii) PPO-driven ABR, and (iii) a DDPG-based continuous-action cache controller. Throughput prediction enables proactive streaming decisions, and the continuous-action cache agent enables smooth, fine-grained control of segment admission, retention, promotion, and eviction across the full lifecycle.

Evaluations using MovieLens and Ghent 4G traces demonstrate improvements in byte hit rate, backhaul traffic, and delayed startup rate, confirming the benefit of predictive, cross-layer DRL orchestration in dynamic edge environments.

The remainder of this paper is organized as follows: Section II reviews related work, Section III details the system design, Section IV presents results, and the Conclusion section concludes.

*Behzad Akbari is the corresponding author:b.akbari@modares.ac.ir*

## II. Literature Review

Video traffic dominates mobile networks, making efficient edge caching essential for reducing latency and sustaining QoE. However, most existing studies address cache placement, admission, and eviction in isolation, and often focus on a single stage of the video life cycle, leaving integrated and lifecycle-aware strategies insufficiently explored.

### A. Video Segment Caching

Segment-based caching aims to improve delivery efficiency and user experience by storing independently retrievable video chunks at edge nodes. Preference-aware methods such as FlyCache [1] and adaptive prefetching schemes [2], [3] highlight the benefits of prediction and personalization. Complementary efforts investigate transcoding-aware caching [4] and cooperative MEC caching [5], emphasizing resource constraints and dynamic user demand.

### B. Cache Placement

Placement strategies increasingly rely on learning-based and adaptive optimization. Multi-agent RL [6], probabilistic RL placement [7], and bandit-based methods [8] demonstrate robustness under uncertain popularity and mobility. Additional studies focus on heterogeneous D2D environments [9] and classic replacement policies [10], underscoring the importance of cross-layer, resource-aware designs.

### C. Edge-Assisted ABR Streaming

Research integrating ABR with edge computing spans compression-aware systems [11], DRL-based adaptation [12], [13], and cross-layer control mechanisms [14], [15]. Cloud–edge cooperation further enhances delivery efficiency [16], [17]. Complementary work explores semantic, or task-driven caching [18]–[20]. Recent studies also examine the role of chunk time in TCP-based adaptive streaming and its impact on QoE [21], reinforcing the need for joint consideration of bitrate control, chunk design, and cache management.

## III. Methodology

### A. Problem Statement

Although recommendation-driven edge caching improves QoE and reduces backhaul usage, existing systems—most notably FlyCache—remain reactive and insufficiently adaptive to dynamic network conditions. These frameworks assume quasi-stationary bandwidth and employ fixed, externally defined ABR policies that are decoupled from caching, resulting in three major limitations:

1) Ignoring Bandwidth Variability: Real 4G/5G networks exhibit significant short-term throughput fluctuations, yet no full-lifecycle caching architecture integrates bandwidth prediction. As a result, ABR decisions often mismatch upcoming channel states, causing stalls and ineffective prefetching.
2) Decoupled ABR–Cache Control: Current ABR logic operates without awareness of cache status or prefix availability, missing opportunities for coordinated decisions and leading to redundant backhaul fetches.
3) Static Cache Policies: Existing placement, admission, and eviction heuristics are optimized for static preference profiles and cannot adapt to rapid shifts in user demand or network congestion, limiting scalability in heterogeneous edge environments.

### B. System Architecture

The proposed system extends FlyCache [1] with a fully integrated, data-driven control framework spanning the entire video streaming pipeline. The architecture consists of three layers, illustrated in Fig. 1.

**User Layer.** Each user follows a Markovian state machine (IDLE, RECO/SEARCH, WATCH, PAUSE, EXIT). Users maintain a playback buffer, a 19-dimensional preference vector updated via exponential moving average (EMA), and throughput histories derived from real 4G traces.

**Edge Layer.** The edge server manages four cache regions—Static Buffer (prefix segments), Maintain Space (full videos), Dynamic Buffer (per-session prefetching), and Victim Space (progressively evicted segments). Three learning modules drive the adaptive behavior: (i) a Transformer-based bandwidth predictor forecasting throughput at +4, +8, and +12 seconds; (ii) a PPO-based ABR agent selecting bitrates from buffer, bandwidth forecasts, and playback features; and (iii) a DDPG-based cache agent operating on a 220-dimensional global state to output continuous joint admission/eviction actions. A lightweight recommendation module ranks candidate videos using cosine similarity while promoting cached content to increase cache hit probability.

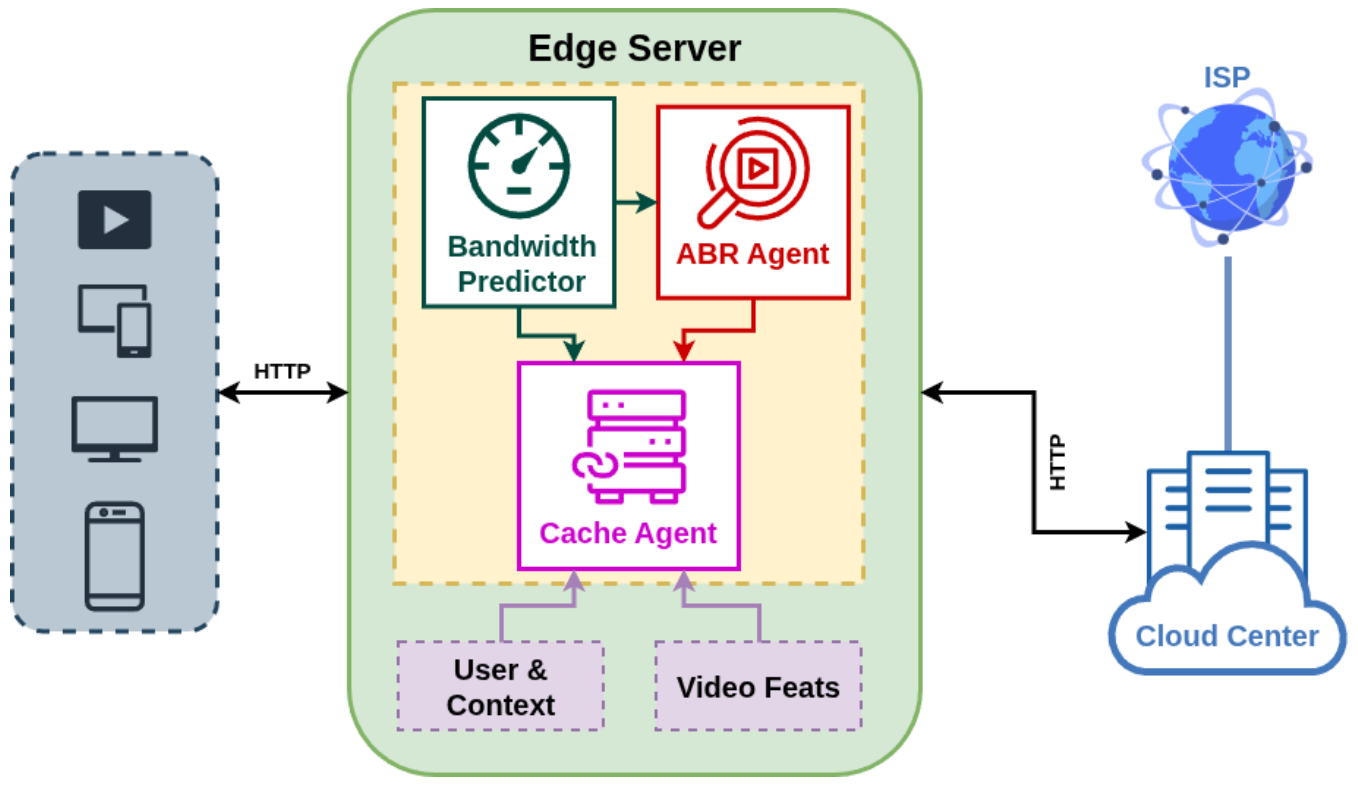

Fig. 1: System Model

TABLE I: Notation Table

| Symbol | Meaning |
|---|---|
| **Users, Videos, and Preferences** | |
| $u$, $v$ | User / video index |
| $\mathcal{S}_v$ | Segment set of video $v$ |
| $\mathbf{f}_v$ | Feature vector of video $v$ |
| $\mathbf{p}_t^u$ | User preference vector at time $t$ |
| $\rho_t^{u,v}$ | Playback progress for user $u$ on $v$ |
| **Network and ABR Variables** | |
| $\theta_u(t)$ | Measured throughput |
| $\mathcal{H}_t^u$ | Throughput history window |
| $\hat{\boldsymbol{\theta}}_t^u$ | Predicted bandwidth (Transformer) |
| $a_t^u$ | ABR action selected by PPO |
| $r_t^u$ | ABR reward |
| **Cache Control Variables** | |
| $\bar{\mathbf{p}}$ | Mean preference of active users |
| $\bar{\boldsymbol{\theta}}$ | Mean predicted bandwidth |
| $\mathbf{a}_t$ | Continuous cache action (DDPG) |
| $R_t$ | Cache reward |
| **Simulation Parameters** | |
| $T$ | Simulation horizon |
| $C$ | Cache capacity (segments) |
| $S$ | Static buffer size |

**Cloud Layer.** The cloud stores the complete video library and performs periodic global preference aggregation. Only essential metadata (updated preference vectors, miss-triggered segment requests) is exchanged with the edge. Together, the components form a closed-loop pipeline: bandwidth prediction guides ABR actions, ABR outcomes and user behavior shape cache decisions, and hit/miss events generate QoE-driven rewards for the DRL agents. The system operates over discrete time slots $t \in \{1, \ldots, T\}$, each corresponding to a 4-second video segment. Let $|\mathcal{U}| = U$ denote the number of users, $|\mathcal{V}| = V$ the number of videos, and $S_v = \{s_{v,0}, \ldots, s_{v,N_v-1}\}$ the segment set of each video $v$. Videos are encoded at discrete bitrates: $B = \{800, 1500, 3000, 6000\}$Kbps. Each user maintains a preference vector updated via an exponential moving average(EMA):

$$\mathbf{p}_t^u = \alpha \cdot \frac{\mathbf{f}_v \rho_t^{u,v}}{\|\mathbf{f}_v \rho_t^{u,v}\|_1} + (1-\alpha)\mathbf{p}_{t-1}^u \quad (1)$$

The edge cache, with total capacity $C$, is divided into four regions [1]: a Static Buffer for prefixes, a Maintain Space for active full videos, a Dynamic Buffer for session-driven prefetching, and a Victim Space for progressively truncated content.

The joint system action consists of ABR decisions and continuous cache-control outputs:

$$\mathcal{A}_t = \{0,1,2,3\}^{|\mathcal{A}_t^{\text{watch}}|} \times [-1,1]^K \tag{2}$$

where $K = 10$ candidate videos. Each cache action $a_{t,i} \in [-1,1]$ determines eviction, demotion, no-op, admission, or prefetching. Traditional full-lifecycle caching approaches [1] assume stable channels and treat ABR and caching separately, resulting in suboptimal behavior under wireless variability. They neglect (i) bandwidth non-stationarity, (ii) the coupling between bitrate selection and cache hit probability, and (iii) real-time preference drift. We therefore optimize the following objective:

$$\max_{\pi} \frac{1}{T}\sum_{t=1}^{T} \mathbb{E}_\pi\Big[\beta_1\,\text{BHR}_t + \beta_2\big(1-\min(\text{StallRatio}_t,1)\big) - \beta_3\,\frac{\text{Backhaul}_t}{10^6}\Big]. \tag{3}$$

### C. Contributions

We propose ProCAVE, a predictive multi-agent DRL architecture that jointly optimizes bandwidth forecasting, bitrate adaptation, and preference-aware cache decisions. The main contributions are summarized below.

1) **Transformer-Based Bandwidth Prediction.** A lightweight Transformer encoder maps the 20-second throughput history $\mathcal{H}_t^u$ to a 3-step forecast:

$$\widehat{\boldsymbol{\theta}}_t^u = \big[\widehat{\theta}_u(t{+}4), \widehat{\theta}_u(t{+}8), \widehat{\theta}_u(t{+}12)\big]^\top \tag{4}$$

2) **PPO-Based ABR Agent.** The ABR policy observes:

$$\mathbf{s}_t^u = \big[b_{t-1}^u,\ \sigma_{k_{t-1}}^u,\ d_{t-1}^u,\ \text{buffer}_t^u,\ \widehat{\theta}_u(t{+}4),\ \widehat{\theta}_u(t{+}8),\ \widehat{\theta}_u(t{+}12),\ \text{stall}_t^u,\ \text{comp}_t^u\big]^\top \tag{5}$$

and selects a bitrate that maximizes long-term QoE.

3) **DDPG-Based Cache Agent.** The cache controller operates on a 220-dimensional global state and outputs continuous actions $\mathbf{a}_t = \mu_\psi(\mathbf{Z}_t) \in [-1,1]^{10}$.

$$\mathbf{Z}_t = [\bar{\mathbf{p}},\ \bar{\boldsymbol{\theta}},\ \mathbf{u}_{\text{util}},\ b_{\text{avg}},\ \text{StallRisk}_t,\ \bigoplus_{i=1}^{10}\mathbf{f}_{v_i}] \tag{6}$$

Soft target updates follow:

$$\phi' \leftarrow \tau\phi + (1-\tau)\phi', \qquad \psi' \leftarrow \tau\psi + (1-\tau)\psi' \tag{7}$$

4) **Lifecycle-Aware Integration.** The agents coordinate across the entire streaming lifecycle:
   - *Before playback:* bandwidth-aware prefix placement in $\mathcal{C}_{\text{stat}}$;
   - *During playback:* ABR-driven promotion and admission in $\mathcal{C}_{\text{main}}$;
   - *After playback:* adaptive truncation in $\mathcal{C}_{\text{vic}}$ based on preference drift and stall risk.

**Algorithm 1** ProCAVE Multi-Agent Control Loop

**Require:** Users $\{u\}$, cache $\mathcal{C}$, models $\mathcal{T}_\phi$, $\pi_\theta$, $\mu_\psi$
1: **for** $t = 1$ to $T$ **do**
2:   **for** each active user $u$ **do**
3:     $\widehat{\boldsymbol{\theta}}_t^u \leftarrow \mathcal{T}_\phi(\mathcal{H}_t^u)$
4:     $a_t^u \sim \pi_\theta(\cdot \mid \mathbf{s}_t^u)$
5:     Fetch segment; update $\mathbf{p}_t^u$ and store $(\mathbf{s}_t^u, a_t^u, r_t^u)$
6:   **end for**
7:   Form global state $\mathbf{Z}_t$
8:   $\mathbf{a}_t \leftarrow \mu_\psi(\mathbf{Z}_t)$
9:   Apply cache operations based on $\mathbf{a}_t$
10:   Observe $R_t$ and store $(\mathbf{Z}_t, \mathbf{a}_t, R_t)$
11:   **if** training **then**
12:     Update $\pi_\theta$ (PPO); update $\mu_\psi$ (DDPG)
13:   **end if**
14: **end for**

**Algorithm 2** Transformer Bandwidth Predictor $\mathcal{T}_\phi$

**Require:** History $\mathcal{H}_t$
1: $\mathbf{X} \leftarrow \text{PosEnc}(\mathcal{H}_t)$
2: **for** 1 to $L$ **do**
3:   $\mathbf{X} \leftarrow \text{LN}(\mathbf{X} + \text{MHSA}(\mathbf{X}))$
4:   $\mathbf{X} \leftarrow \text{LN}(\mathbf{X} + \text{FFN}(\mathbf{X}))$
5: **end for**
6: $\mathbf{z} \leftarrow \text{MeanPool}(\mathbf{X})$
7: **return** $\widehat{\boldsymbol{\theta}}_t = W_o\mathbf{z} + b_o$

### D. Multi-Agent Control Architecture

ProCAVE employs a three-agent framework for joint optimization of video streaming under dynamic network conditions: (i) a Transformer-based agent predicting future bandwidth from temporal network statistics; (ii) a PPO-based agent selecting bitrates using real-time measurements and bandwidth forecasts; and (iii) a DDPG-based agent managing edge cache (segment storage/eviction) to maximize long-term QoE.

## IV. Experimental Results and Analysis

We evaluate ProCAVE on two real-world datasets capturing user preference dynamics and wireless variability.

*a) MovieLens (ml-latest-small):* We use the *MovieLens-latest-small* dataset [22] (610 users, 9,742 movies, 100,836 ratings). Movies are represented by 19-dimensional genre vectors [1]; user preferences are initialized via rating-weighted averaging and updated online with an exponential moving average.

*b) Ghent 4G Throughput Traces:* Bandwidth dynamics are modeled using 40 real-world 4G/LTE traces from [23] ($\approx 1$ Hz sampling, up to 95 Mbps). A 531 s representative trace yields 510 overlapping samples (10 s input windows; 3-step forecasts at +4, +8, +12 s), used to train the Transformer predictor (input: (10,1); output: future bandwidth estimates).

Together, these datasets enable evaluation under heterogeneous, evolving user interests and realistic short-term network

**Algorithm 3** PPO-Based ABR Update

**Require:** State $\mathbf{s}_t$, old policy $\pi_{\theta_{old}}$, advantage $\hat{A}_t$

1: $a_t \sim \pi_\theta(\cdot \mid \mathbf{s}_t)$
2: $r_t = \frac{\pi_\theta(a_t)}{\pi_{\theta_{old}}(a_t)}$
3: $\tilde{r}_t = \text{clip}(r_t, 1-\epsilon, 1+\epsilon)$
4: $\mathcal{L} = \min(r_t\hat{A}_t, \tilde{r}_t\hat{A}_t)$
5: Update $\theta$ to maximize $\mathcal{L}$ + entropy

**Algorithm 4** DDPG Cache Update

**Require:** State $\mathbf{Z}_t$, actor $\mu_\psi$, critic $Q_\phi$

1: $\mathbf{a}_t \leftarrow \mu_\psi(\mathbf{Z}_t) + \mathcal{N}$
2: Store transition $(\mathbf{Z}_t, \mathbf{a}_t, R_t, \mathbf{Z}_{t+1})$
3: **if** update **then**
4: Sample minibatch from buffer
5: $y_i = R_i + \gamma Q_{\phi'}(\mathbf{Z}_{i+1}, \mu_{\psi'}(\mathbf{Z}_{i+1}))$
6: Update critic using $(y_i - Q_\phi)^2$
7: Update actor using $\nabla_\psi Q_\phi(\mathbf{Z}_i, \mu_\psi(\mathbf{Z}_i))$
8: Soft-update targets
9: **end if**

dynamics—reflecting full-lifecycle caching challenges in dynamic edge environments.

The ProCAVE multi-agent edge caching framework was implemented in Python 3.12.8 and evaluated on a workstation with an Intel Core i5-12400F, 32 GiB RAM, and NVIDIA RTX 3060 Ti (8 GiB VRAM). As shown in Figure 2, over 250 time steps, ProCAVE significantly outperforms FlyCache across all key metrics: Byte Hit Rate (BHR) increases from ∼0.45 to ∼0.70, backhaul traffic reduces from $1.4 \times 10^8$ to $1.1 \times 10^8$ bytes, and Delayed Startup Rate (DSR) declines from 0.60 to 0.45—demonstrating superior demand prediction, resource efficiency, and QoE under dynamic workloads.

Figure 3 confirms ProCAVE's scalability: BHR rises monotonically to ∼0.34 at 350 segments (over 10× FlyCache's ∼0.03), while FlyCache peaks early and degrades. ProCAVE reduces backhaul traffic by up to 38.5% at 300 segments without instability, whereas FlyCache exhibits anomalous spikes. DSR improves steadily to ∼0.60 (vs. FlyCache's ∼1.35 at 300 segments), indicating robust startup performance. These results validate ProCAVE's capacity for joint optimization of infrastructure efficiency and user experience, overcoming baseline scalability limitations and establishing its viability for large-scale edge deployments.

## Conclusion

We propose ProCAVE, a predictive multi-agent DRL framework that jointly optimizes bandwidth forecasting, ABR control, and preference-aware caching across the video lifecycle. By integrating a lightweight Transformer for throughput prediction, a PPO-based ABR policy, and a DDPG-driven continuous-action cache controller, ProCAVE enables proactive, cross-layer adaptation, overcoming the decoupled, reactive design of prior systems like FlyCache.

TABLE II: Simulation and Learning Parameters

| Parameter | Value |
|---|---|
| **DDPG Cache Agent** | |
| Actor learning rate | $1 \times 10^{-4}$ |
| Critic learning rate | $1 \times 10^{-3}$ |
| Discount factor $\gamma$ | 0.99 |
| Soft update rate $\tau$ | 0.005 |
| Replay buffer size | $10^6$ |
| **PPO ABR Agent** | |
| Policy learning rate | $3 \times 10^{-4}$ |
| Discount factor $\gamma$ | 0.99 |
| Clip ratio | 0.20 |
| Observation window | 5 segments |
| **Simulation Settings** | |
| Horizon $T$ | 250 steps |
| Cache capacity $C$ | 100–500 segments |
| Static buffer $S$ | $0.2C$ |
| Bitrate levels | 6 levels (300–4300 kbps) |

Evaluated on real MovieLens and Ghent 4G traces, ProCAVE improves byte hit rate by $\sim 56\%$, reduces backhaul traffic by 21%, and lowers delayed startup rate by 25% over 250 steps. Under cache scaling (100–350 segments), it achieves monotonic BHR growth (up to ∼0.34), outperforming FlyCache by over an order of magnitude at high capacity while avoiding traffic spikes and latency degradation. These results confirm that predictive, DRL-based orchestration significantly enhances both system efficiency and QoE in dynamic edge environments. Future work includes federated deployment and integration with semantic video understanding.

## References

[1] S. Cao, Q. Zheng, Z. Zhan, Y. Yang, H. Lv, D. Zheng, and W. Zhang, "Flycache: Recommendation-driven edge caching architecture for full life cycle of video streaming," *Digital Communications and Networks*, vol. 11, no. 4, pp. 961–974, 2025. [Online]. Available: https://www.sciencedirect.com/science/article/pii/S235286482500001X

[2] J. Aguilar-Armijo, C. Timmerer, and H. Hellwagner, "Segment prefetching at the edge for adaptive video streaming," in *2022 18th International Conference on Wireless and Mobile Computing, Networking and Communications (WiMob)*, 2022, pp. 339–344.

[3] ——, "Space: Segment prefetching and caching at the edge for adaptive video streaming," *IEEE Access*, vol. 11, pp. 21 783–21 798, 2023.

[4] A. Erfanian, H. Amirpour, F. Tashtarian, C. Timmerer, and H. Hellwagner, "Cd-lwte: Cost- and delay-aware light-weight transcoding at the edge," *IEEE Transactions on Network and Service Management*, vol. 20, no. 3, pp. 3104–3118, 2023.

[5] A. Lekharu, P. Gupta, A. Sur, and M. Patra, "Collaborative video caching in the edge network using deep reinforcement learning," *ACM Trans. Internet Things*, vol. 5, no. 3, Jun. 2024. [Online]. Available: https://doi.org/10.1145/3664613

[6] T. Zhang, X. Fang, Z. Wang, Y. Liu, and A. Nallanathan, "Stochastic game based cooperative alternating q-learning caching in dynamic d2d networks," *IEEE Transactions on Vehicular Technology*, vol. 70, no. 12, pp. 13 255–13 269, 2021.

[7] M. Amidzadeh, H. Al-Tous, O. Tirkkonen, and J. Zhang, "Joint cache placement and delivery design using reinforcement learning for cellular networks," in *2021 IEEE 93rd Vehicular Technology Conference (VTC2021-Spring)*, 2021, pp. 1–6.

[8] Y. Han, L. Ai, R. Wang, J. Wu, D. Liu, and H. Ren, "Cache placement optimization in mobile edge computing networks with unaware environment—an extended multi-armed bandit approach," *IEEE Transactions on Wireless Communications*, vol. 20, no. 12, pp. 8119–8133, 2021.

[9] N. Abdolkhani, M. Eslami, J. Haghighat, and W. Hamouda, "Optimal caching policy for d2d assisted cellular networks with different cache size devices," *IEEE Access*, vol. 10, pp. 99 353–99 360, 2022.

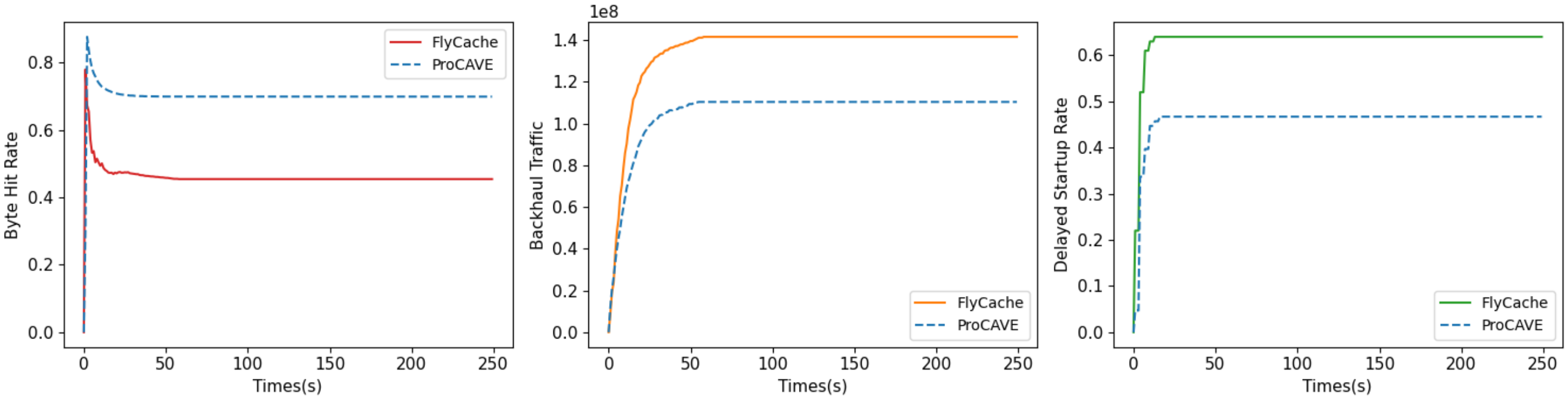

Fig. 2: The Trend of Byte Hit Rate, Backhaul Traffic, and Delayed Startup Rate Over Time

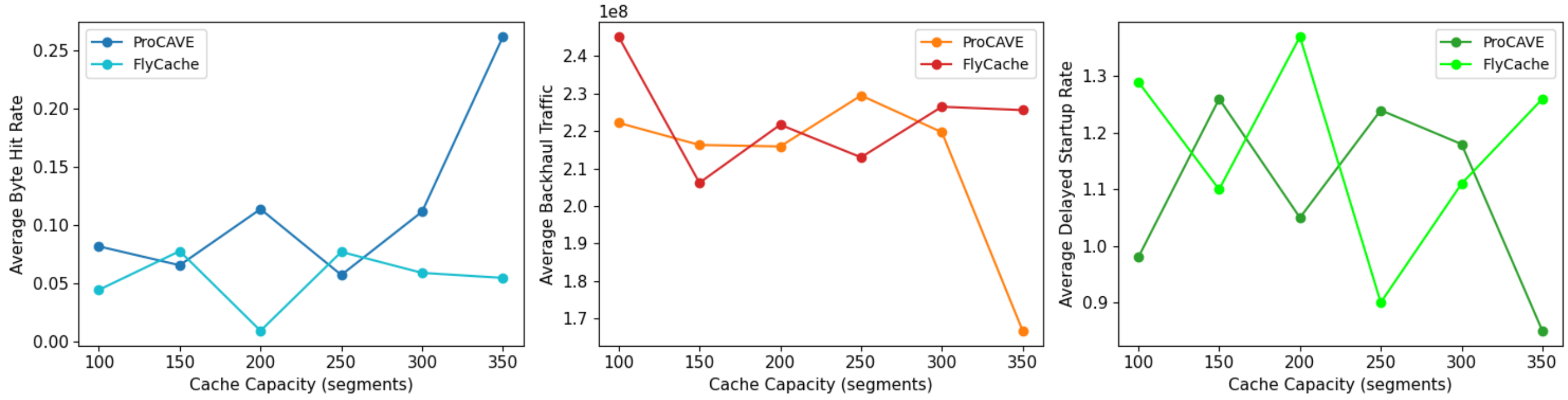

Fig. 3: Impact of Cache Capacity on Byte Hit Rate, Backhaul Traffic, and Delayed Startup Rate

[10] F. Zharfan, L. D. Hasnaa, R. Muldina Negara, and N. R. Syambas, "Comparison of caching replacement policies in changing the number of interest packets on named data networks using mininet-ndn," in *2021 15th International Conference on Telecommunication Systems, Services, and Applications (TSSA)*, 2021, pp. 1–8.

[11] A. Hojjat, J. Haberer, and O. Landsiedel, "Mcucoder: Adaptive bitrate learned video compression for iot devices," 2024. [Online]. Available: https://arxiv.org/abs/2411.19442

[12] S. Bi, H. Chen, X. Li, S. Wang, Y. Wu, and L. Qian, "A two-stage deep reinforcement learning framework for mec-enabled adaptive 360-degree video streaming," *IEEE Transactions on Mobile Computing*, vol. 23, no. 12, pp. 14 313–14 329, 2024.

[13] D. Zhang, L. Wei, K. Shen, H. Zhu, D. Wang, and F. Wang, "Trimstream: Adaptive realtime video streaming through intelligent frame retrospection in adverse network conditions," *IEEE Transactions on Mobile Computing*, vol. 23, no. 12, pp. 11 240–11 252, 2024.

[14] Y. F. Yeznabad, M. Helfert, and G.-M. Muntean, "Qoe-driven cross-layer bitrate allocation approach for mec-supported adaptive video streaming," *IEEE Transactions on Network and Service Management*, vol. 21, no. 6, pp. 6857–6874, 2024.

[15] X. Xiao, Y. Zuo, M. Yan, W. Wang, J. He, and Q. Zhang, "Task-oriented video compressive streaming for real-time semantic segmentation," *IEEE Transactions on Mobile Computing*, vol. 23, no. 12, pp. 14 396–14 413, 2024.

[16] H. Zhan, L. Fan, C. Li, X. Lei, and F. Li, "Cloud–edge learning for adaptive video streaming in b5g internet of things systems," *IEEE Internet of Things Journal*, vol. 11, no. 24, pp. 40 140–40 148, 2024.

[17] A. Xiao, S. Wu, Y. Ou, N. Chen, C. Jiang, and W. Zhang, "Qoe-fairness-aware bandwidth allocation design for mec-assisted abr video transmission," *IEEE Transactions on Network and Service Management*, vol. 22, no. 1, pp. 499–515, 2025.

[18] J. Zheng, T. H. Luan, G. Li, Z. Yin, Y. Wu, and M. Dong, "Acdv: Adaptive content delivery for vehicular digital twin networks," *IEEE Transactions on Vehicular Technology*, vol. 74, no. 5, pp. 7084–7098, 2025.

[19] Y. F. Yeznabad, M. Helfert, and G.-M. Muntean, "Enhancing qoe through adaptive bitrate allocation in collaborative mec-enabled wireless networks," *IEEE Transactions on Vehicular Technology*, vol. 74, no. 6, pp. 9491–9505, 2025.

[20] Y. Wu, T.-T. Nguyen, L. Liebe, Q. Tau, P. E. Campos, J. Cheng, and D. Lee, "How2compress: Scalable and efficient edge video analytics via adaptive granular video compression," 2025. [Online]. Available: https://arxiv.org/abs/2510.18409

[21] Y. Chatri, B. Akbari, and P. Goudarzi, "Investigating the impact of chunk time on video streaming performance in tcp-based networks," in *2025 29th International Computer Conference, Computer Society of Iran (CSICC)*, 2025, pp. 1–5.

[22] F. M. Harper and J. A. Konstan, "The movielens datasets: History and context," *ACM Trans. Interact. Intell. Syst.*, vol. 5, no. 4, Dec. 2015. [Online]. Available: https://doi.org/10.1145/2827872

[23] J. van der Hooft, S. Petrangeli, T. Wauters, R. Huysegems, P. Rondao Alface, T. Bostoen, and F. De Turck, "HTTP/2-based adaptive streaming of HEVC video over 4G/LTE networks," *IEEE Communications Letters*, vol. 20, no. 11, pp. 2177–2180, 2016.